\documentclass[aps,prl,showpacs,showkeys,nofootinbib]{revtex4-1}
\usepackage{amssymb}
\usepackage{amsbsy}         % возможность использования \boldsymbol для греческих
\usepackage[T2A,OT1]{fontenc}
\usepackage[cp1251]{inputenc}         % Кодировка документа
\usepackage[russian,ukrainian,english]{babel} % Многоязычность документа
\usepackage{graphicx} %for figures
\usepackage{latexsym}
\usepackage{amsfonts}
\usepackage{boxedminipage}

  \usepackage{xcolor}[2007/01/21]
\usepackage[dvipdfm,bookmarks=true,bookmarksnumbered,bookmarksopen]{hyperref}
\hypersetup{pdfpagemode=FullScreen,colorlinks=true,
citecolor=magenta,unicode,bookmarksopenlevel=3,
pdftoolbar=true,pdfauthor={Helen
Gomonay},pdftitle={Current-induced dynamics in antiferromagnetic
texture}}
\begin{document}
\selectlanguage{english} \sloppy
\title{Hydrodynamic theory of coupled current and magnetization dynamics in spin-textured antiferromagnets}
\author{H.~Gomonay and       V.~Loktev}
\affiliation{National Technical University of Ukraine ``KPI'', 37
ave Peremogy, 03056, Kyiv, Ukraine, Bogolyubov Institute for
Theoretical Physics National Academy of Sciences of Ukraine, 14-b
Metrologichna Str., 03680, Kyiv, Ukraine}
\begin{abstract}
Antiferromagnets with vanishingly small (or zero) magnetization
are interesting candidates for spintronics applications. In the
present paper we propose two models for description of the
current-induced phenomena in antiferromagnetic textures. We show
that the magnetization that originates from rotation or
oscillations of antiferromagnetic vector can, via $sd$-exchange
coupling, polarize the current and give rise to adiabatic and
nonadiabatic spin torques. Due to the Lorentz-type dynamics of
antiferromagnetic moments (unlike the Galilenian-like dynamics in
ferromagnets), the adiabatic spin torque affects the
characteristic lengthscale of the moving texture. Nonadiabatic
spin torque contributes to the energy pumping and can induce the
stable motion of antiferromagnetic texture, but, in contrast to
ferromagnets, has pure dynamic origin. We also consider the
current-induced phenomena in artificial antiferromagnets where the
current maps the staggered magnetization of the structure. In this
case the effect of nonadiabatic spin torque is similar to that in
ferromagnetic constituents of the structure. In particular, the
current can remove degeneracy of the translational
antiferromagnetic domains indistinguishable in the external
magnetic field and thus can set into motion  the 180$^\circ$
domain wall.
\end{abstract}
\maketitle

\section{Introduction}
Spin-polarized current flowing through the magnetic layers gives
rise to variety of different phenomena with interesting physics
and wide range of applications. In particular, the current can
transfer a torque and thus produce a spin-motive force for the
system of localized magnetic moments. This effect is responsible
for rotation of ferromagnetic magnetization in discrete systems or
movement of domain wall in continuous textures.

Spin-torque phenomena, as predicted by Slonczewski
\cite{Slonczewski:1996}
 and Berger \cite{Berger:1996}, originate
from the spin-conserving exchange interactions between itinerant
and localized electrons (so-called $sd$-exchange) and are usually
modeled with the Heisenberg-type
 Hamiltonian
 \begin{equation}\label{eq_exchange_hamiltonian}
    \hat\mathcal{H}_{\mathrm{sd}}=-\sum_{\mathbf{R}_\mathbf{n}}J_{\mathrm{sd}}(\mathbf{r}-\mathbf{R}_\mathbf{n})\hat{\mathbf{s}}(\mathbf{r})\hat{\mathbf{S}}(\mathbf{R}_\mathbf{n})
\end{equation}
 where $J_{\mathrm{sd}}$ is the exchange integral,
 $\hat{\mathbf{s}}(\mathbf{r})$ and
 $\hat{\mathbf{S}}(\mathbf{R}_\mathbf{n})$ are spin operators  of
 free ($s$) and localized ($d$) at a site
 $\mathbf{R}_\mathbf{n}$ electrons, respectively.

In the semiclassical approach \cite{Zhang:PhysRevLett.93.127204}
that assumes rather slow dynamics of localized spins compared to
the itinerant ones, the system of localized magnetic moments is
treated classically. Particularly, in ferromagnets (FMs) the spin
operators $\hat{\mathbf{S}}(\mathbf{R}_\mathbf{n})$ are  replaced
with the magnetization vector $\mathbf{M}(\mathbf{r},t)$, and the
averaged over all electronic states spin $\hat{\mathbf{s}}$ is
replaced with the normalized electron magnetization density
$\mathbf{m}$ ($|\mathbf{m}|=1$). Thus, the energy density of
$sd$-exchange, derived from Hamiltonian
(\ref{eq_exchange_hamiltonian}), takes a form
 \begin{equation}\label{eq_Heisenberg-like term}
    \mathcal{H}_{\mathrm{sd}}(\mathbf{r})=-\frac{J_{\mathrm{sd}}}{M_s}\mathbf{m}\cdot\mathbf{M},
\end{equation}
where $M_s$ is saturation magnetization.

 Interactions described by Eqs.~(\ref{eq_exchange_hamiltonian})
 and (\ref{eq_Heisenberg-like term}) determine the processes of
 current polarization and transfer of spin torque. In FMs with the
 pronounced value of equilibrium magnetization $\mathbf{M}$ the
 effects related with
 $sd$-interaction are quite strong.

 However, application of the same ideas to other magnetic
 materials, -- antiferromagnets (AFMs), with zero or vanishingly
 small equilibrium magnetization, -- poses new challenges as
 compared to FMs. Macroscopic magnetization $\mathbf{M}_\mathrm{AFM}$ in AFMs has mainly
 dynamic origin. Its value is weakened (compared to FMs) by
 relatively small magnetic susceptibility. On the other hand,
AFM ordering is characterized with macroscopic vector(s)
$\mathbf{L}_k$ ($k=1,2\ldots N_\mathrm{sub}$, where
$N_\mathrm{sub}$ depends upon the number of magnetic sublattices)
which
 reproduce a space distribution of staggered magnetization and in
 most cases belong to another than $\mathbf{M}_\mathrm{AFM}$ irreducible
 representation of the space group (i.e., have different symmetry
 properties). In this case the semiclassical expression for the
 density of $sd$-exchange, as follows from general symmetry
 considerations, takes a
 form (cp. with (\ref{eq_Heisenberg-like term})):
 \begin{equation}\label{eq_sd interaction_AFM}
 \mathcal{H}_{\mathrm{sd}}(\mathbf{r})=-\frac{1}{{M_s}}\left[J^{\mathrm{FM}}_{\mathrm{sd}}\mathbf{m}\cdot\mathbf{M}_\mathrm{AFM}+J^{\mathrm{AFM}}_{\mathrm{sd}}\sum_{k=1}^{N_\mathrm{sub}}\mathbf{l}_k\cdot\mathbf{L}_k\right],
\end{equation}
where vectors $\mathbf{l}_k$, which characterize the itinerant
electrons, should belong to the same irreducible representation as
$\mathbf{L}_k$ and $J^{\mathrm{FM}}_{\mathrm{sd}}$,
$J^{\mathrm{AFM}}_{\mathrm{sd}}$ are the corresponding exchange
constants.

The first, ``standard'', or FM-like, term in the Eq.~(\ref{eq_sd
interaction_AFM}) may induce rotation of AFM moments once the
current is already polarized by the external FM layer
\cite{Haney:2007(2),Tsoi:2007,gomo:2010,
Gomonay:PhysRevB.85.134446}. The second, ``specific'', or AFM-like
term, if any, may induce a special -- AFM -- polarization of
current and thus should be important in AFM textures (see, e.g.,
Ref.~\cite{Nunez:2005}). Whether both terms exist and how
important they are significantly depends upon topology of the
Fermi surface and density of states.

In the present paper we address two mutually related questions:
 \emph{i}) what current-induced effects could be anticipated in
 AFM textures from the first ($J^{\mathrm{FM}}_{\mathrm{sd}}\ne0,
 J^{\mathrm{AFM}}_{\mathrm{sd}}=0$) and the second ( $J^{\mathrm{AFM}}_{\mathrm{sd}}\ne0$) constituents of $sd$-exchange; and \emph{ii}) in what cases spin density of the
 itinerant (represented by AFM vectors $\mathbf{l}_k$)
 electrons may have space modulation that identically maps the space
 distribution of localized moments imposed by crystal lattice.
 Starting from hydrodynamic approach for microscopic dynamics of
 AFMs
 \cite{Andreev:1980} we derive the close set of equations for AFM vectors
in the presence of current assuming the FM-like
($J^{\mathrm{AFM}}_{\mathrm{sd}}=0$) form of $sd$-exchange. We
show existence of adiabatic and nonadiabatic spin torques (AST and
NAST, respectively) and demonstrate their dynamic origin. NAST,
though small compared with AST (and NAST in ferromagnetic
textures)
 can compensate the internal losses in the moving domain
walls.

To illustrate the role of the second, AFM-like contribution to
$sd$-exchange, we consider the current-induced dynamics in
artificial AFMs. AST and NAST in this case have the same values
and reveal themselves in a similar way as in FMs. In particular,
the current (in contrast to constant magnetic field ) can set into
motion the 180$^\circ$ AFM domain wall (DW) that separates
translational domains.

\section{Effect of ferromagnetic-like $sd$-exchange: antiferromagnetic texture}
Let us consider an AFM conductor with well separated systems of
localized and free (conduction) electrons. Good example of such
materials is given by AFM metals FeMn\footnote{~Though FeMn is
frequently considered as itinerant AFM (see e.g.
\cite{Endoh:1973}), its electronic and magnetic structure is still
unclear and can be treated as consisting of localized and
conduction electrons, especially in thin films \cite{Kuch:2004}.}
and IrMn, or by metallic antiperovskites Mn$_3$MN (where M=Ag,
Ni). Following the phenomenological approaches deduced for FM
textures (see, e.g. Refs. \cite{Zhang:PhysRevLett.93.127204,
Sun:2011EPJB_79_449S}) we describe the average magnetization of
free electrons with the field variable $\mathbf{m}(\mathbf{r},t)$
and assume $sd$-exchange interaction in a form (\ref{eq_sd
interaction_AFM}) with $J^{\mathrm{AFM}}_{\mathrm{sd}}=0$.

Description of localized moments needs some special comments. In
spite of diversity of possible magnetic structures, the
low-frequency dynamics of AFMs can be expressed in terms of at
most three independent variables that represent a so-called
``solid-like'' rotation of localized magnetic moments
\cite{Andreev:1980}. In this section we consider the case of
multisublattice AFM with isotropic magnetic susceptibility $\chi$
and parametrize spin rotations with the Gibbs' vector
$\boldsymbol\varphi=\varphi \mathbf{e}$, where the unit vector
$\mathbf{e}$ defines an instantaneous rotation axis,
$\varphi=\tan(\theta/2)$, and $\theta$ is the rotation angle (for
the details see the papers Refs.~\cite{Andreev:1980,
Gomonay:PhysRevB.85.134446}). While the components of Gibbs'
vector $\boldsymbol\varphi$ form a set of coordinates, the
components of spin rotation frequency, expressed as follows:
\begin{equation}\label{eq_rotation_frequency}
    \mathbf{\Omega}=2\frac{\dot{\boldsymbol\varphi}+\boldsymbol\varphi\times\dot{\boldsymbol\varphi}}{1+{\boldsymbol\varphi^2}},
\end{equation}
are conjugated generalized velocities. Here the sign ``$\times$''
means cross-product.

Then, macroscopic magnetization of AFM is proportional to spin
rotation frequency, $\mathbf{\Omega}$, and external magnetic field
$\mathbf{H}$:
\begin{equation}\label{eq_magnetization_1}
   \mathbf{M}_{\mathrm{AFM}}=\frac{\chi}{\gamma}\left(\boldsymbol\Omega+\gamma
   \mathbf{H}\right),
\end{equation}
where $\gamma$ is gyromagnetic ratio.

The dynamic equations for localized AFM moments are deduced from
the spin conservation principle\footnote{~Strictly speaking, in
the materials with the pronounced spin-orbit coupling one should
start from the conservation law for the total angular momentum.
However, for the sake of simplicity, we exclude spin-lattice
interactions.} which takes a form of the balance equation:
\begin{equation}\label{eq_dynamics_2}
   \frac{d\mathbf{M}_\mathrm{AFM}}{dt}=\nabla\cdot
\hat{\boldsymbol{\Pi}},\quad \textrm{or}\quad
\frac{dM_{\mathrm{AFM}}^{(\alpha)}}{dt}=\frac{\partial\Pi_{\alpha\beta}}{\partial x_\beta}, %%sign--?
\end{equation}
where $\hat{\boldsymbol{\Pi}}$ is the 2-nd rank tensor of the
magnetization flux density induced, in particular, by
spin-polarized current. With account of the relation
(\ref{eq_magnetization_1}) the dynamic Eq.~(\ref{eq_dynamics_2})
for the localized AFM moments in the most general case can be
written as follows:
\begin{equation}\label{eq_dynamic_5}
   \frac{2\chi}{\gamma^2}\lambda_{\beta\alpha}\frac{d}{dt}\left[\left(\Omega_\beta+\gamma
    H_\beta\right)\right]+\frac{2\chi}{\gamma}\lambda_{\beta\alpha}(\mathbf{H}\times\boldsymbol{\Omega})_\beta+\frac{\partial
    U_{\mathrm{AFM}}}{\partial
    \varphi_\alpha}=\frac{2}{\gamma}\lambda_{\beta\alpha}\nabla_\delta\Pi_{\beta\delta},
\end{equation}
where the potential $U_\mathrm{AFM}(\boldsymbol{\varphi})$
describes the magnetic anisotropy of AFM and includes gradient
terms of the exchange nature, tensor $\lambda_{\alpha\beta}$
defines the metrics in $\boldsymbol{\varphi}$ space:
\begin{equation}\label{eq_metric tensor}
  \lambda_{\alpha\beta}=\frac{\delta_{\alpha\beta}+\varepsilon_{\alpha\gamma\beta}\varphi_\gamma}{1+{\boldsymbol\varphi^2}},
\end{equation}
and $\varepsilon_{\alpha\gamma\beta}$ is the (completely
antisymmetric)
 Levi-Civita symbol.

Analysis of the Eq.~(\ref{eq_magnetization_1}) shows that AFMs
with an arbitrary magnetic structure have nonzero magnetization
$\mathbf{M}_{\mathrm{AFM}}$ even in the absence (or neglection) of
the external Oersted magnetic field, though this magnetization
 has pure dynamic origin ($\mathbf{M}_{\mathrm{AFM}}\propto \boldsymbol{\Omega})$.  So, AFM can
polarize the conduction electrons through the $sd$-exchange
interactions (\ref{eq_sd interaction_AFM}) even if
$J^{\mathrm{AFM}}_{\mathrm{sd}}=0$.

To obtain equations for normalized magnetization of free
electrons, $\mathbf{m}(\mathbf{r},t)$, we represent it as a sum of
equilibrium, $\mathbf{m}_\mathrm{eq}$, and small nonequilibrium,
$\delta\mathbf{m}(\mathbf{r},t)$, contributions:
\begin{equation}\label{eq_mangetization_conduction}
\mathbf{m}(\mathbf{r},t)=\mathbf{m}_\mathrm{eq}(\mathbf{r})+\delta\mathbf{m}(\mathbf{r},t).
\end{equation}

If time variation of localized moments is slow compared with the
carrier's spin-flip relaxation, then, equilibrium magnetization of
conduction electrons maps distribution of AFM magnetization
$\mathbf{M}_{\mathrm{AFM}}$. Thus, in analogy with FM,
\begin{equation}\label{eq_current-polarization}
\mathbf{m}_\mathrm{eq}(\mathbf{r})=n_\mathrm{eq}\frac{\mathbf{M}_{\mathrm{AFM}}}{M_s
}=\frac{\chi }{\gamma }\frac{n_\mathrm{eq}}{M_s
}\boldsymbol{\Omega},
\end{equation}
 where $n_\mathrm{eq}$ is the local equilibrium density of carriers whose spin  is
parallel to $\mathbf{M}_\mathrm{AFM}$.

On the other hand, nonequilibrium magnetization $\delta\mathbf{m}$
is created in the AFM  texture due to time and spatial variation
of $\mathbf{M}_\mathrm{AFM}$. For the case of slow space
variations (i.e. the spin-diffusion length is much smaller than
the typical DW width) the dynamic equation for $\delta\mathbf{m}$
takes a form similar to that in FMs
\cite{Zhang:PhysRevLett.93.127204}:
\begin{equation}\label{eq_nonequilibrium_magnetization}
-\frac{\delta\mathbf{m}}{\tau_\mathrm{sf}}-\frac{J_\mathrm{sd}}{\hbar
M_s}\delta\mathbf{m}\times\mathbf{M}_\mathrm{AFM}=\frac{n_\mathrm{eq}}{M_s}\frac{\partial
\mathbf{M}_\mathrm{AFM}}{\partial
t}-\frac{P}{eM_s}(\mathbf{j}\cdot\nabla)\mathbf{M}_\mathrm{AFM}
\end{equation}
where $\tau_\mathrm{sf}$ is the time of spin-flip relaxation,
 $P$ is the spin polarization
factor, $e$ is electron charge, $\mathbf{j}$ is the electric
current density.

The left-hand side of Eq.~(\ref{eq_nonequilibrium_magnetization})
includes two terms: the first one corresponds to spin relaxation
of free electrons with the characteristic time $\tau_\mathrm{sf}$,
and the second one describes rotation of free electron
magnetization around localized magnetization
$\mathbf{M}_\mathrm{AFM}$. In FMs, where
$|\mathbf{M}_\mathrm{FM}|=M_s$, the second term is much greater
than the first one, corresponding relation
$\tau_\mathrm{sf}J_\mathrm{sd}/\hbar\propto 10\div10^2$ is based
on the typical values for FM metals like Fe, Ni and Co:
$\tau_\mathrm{sf}\propto10^{-12}$~s, $J_\mathrm{sd}\propto 1$~eV
\cite{Zhang:PhysRevLett.93.127204}. In contrast, in AFM materials
the relation between these two terms is reversed. Taking for
estimation $\Omega$ of the order of AFMR frequency
$\Omega_\mathrm{AFMR}\propto\gamma
M_s\sqrt{H_\mathrm{an}H_\mathrm{ex}}$ (where $H_\mathrm{an}$ and
$H_\mathrm{ex}= M_s/\chi$ are the
 anisotropy and exchange fields for localized moments) we get from (\ref{eq_magnetization_1}) for
 the typical AFMs (FeMn, IrMn, NiO)
$|\mathbf{M}_\mathrm{AFM}|/M_s\propto\sqrt{H_\mathrm{an}/H_\mathrm{ex}}\propto
1\div3\cdot 10^{-2}$. Typical value of $J_\mathrm{sd}\propto
0.1\div0.01$~eV \cite{Vonsovskii}, so, for the same value of
spin-flip relaxation,
$\tau_\mathrm{sf}J_\mathrm{sd}|\mathbf{M}_\mathrm{AFM}|/(\hbar
M_s)\propto 10^{-2}\ll 1$.%%file data structures.xls, for IrMn

With account of the above relation, the
Eq.~(\ref{eq_nonequilibrium_magnetization}) can be solved in terms
of rotation frequency as follows:
\begin{equation}\label{eq_momequilibrium_spin_density}
  \delta\mathbf{m}=\frac{\chi\tau_\mathrm{sf}}{\gamma M_s}\left[-n_\mathrm{eq}\dot{\boldsymbol{\Omega}}+
  \frac{P}{e}(\mathbf{j}\cdot\nabla)\boldsymbol{\Omega}\right].
\end{equation}

Magnetization of current  enters dynamic Eqs.~(\ref{eq_dynamic_5})
for AFM vectors in two ways. First, equilibrium magnetization
(\ref{eq_current-polarization}) of free electrons produces the
magnetization flux
\begin{equation}\label{eq_spin_flux}
 \boldsymbol{\Pi}=\frac{\chi\mu_B P}{\gamma eM_s}\mathbf{j}\otimes\boldsymbol{\Omega}=\frac{\chi }{\gamma
}b_\mathrm{AFM}\mathbf{j}\otimes\boldsymbol{\Omega},\quad
b_\mathrm{AFM}\equiv\frac{\mu_BP}{eM_s},
\end{equation}
that should be substituted into r.h.s. of
Eq.~(\ref{eq_dynamic_5}), and $\mu_B$ is the Bohr magneton.

Second, nonequilibrium magnetization, due to $sd$-interactions,
produces in AFM an additional magnetic field
\begin{equation}\label{eq_effective_field}
\gamma \mathbf{H}_\mathrm{add}\equiv-\gamma \frac{\partial
\mathcal{H}_{\mathrm{sd}}}{\partial \mathbf{M}_\mathrm{AFM}
}=\gamma \frac{J^\mathrm{FM}_\mathrm{sd}}{M_s}\delta\mathbf{m}%\frac{J^\mathrm{F}_{\mathrm{sd}}\chi\tau_\mathrm{sf}}{\gammaM^2_s}n_\mathrm{eq}\dot{\boldsymbol{\Omega}}
=-\chi\epsilon_\mathrm{AFM}\dot{\boldsymbol{\Omega}}
+c_\mathrm{AFM}(\mathbf{j}\cdot\nabla)\boldsymbol{\Omega},
\end{equation}
where we introduced the phenomenological constants
\begin{equation}\label{eq_constants}
  \epsilon_\mathrm{AFM}\equiv\frac{
\tau_\mathrm{sf}J^\mathrm{FM}_\mathrm{sd}}{M^2_s}n_\mathrm{eq},\qquad
c_\mathrm{AFM}\equiv\frac{ \chi\tau_\mathrm{sf}
}{eM^2_s}J^\mathrm{FM}_\mathrm{sd}P=b_\mathrm{AFM}\frac{\tau_\mathrm{sf}J^\mathrm{FM}_\mathrm{sd}}{\mu_BH_\mathrm{ex}},
\end{equation}
and $H_\mathrm{ex}= M_s/\chi$, as above.

Additional, current-induced  field (\ref{eq_effective_field})
plays a role of the external dissipation force that results, as
will be shown below, in variation (relaxation or pumping) of the
magnetic energy of localized moments.

Substituting (\ref{eq_spin_flux}) and (\ref{eq_effective_field})
 into (\ref{eq_dynamic_5}) one gets
equation for AFM texture in the presence of spin-polarized dc
current:
\begin{eqnarray}\label{eq_dynamic_6}
 \dot{\boldsymbol{\Omega}}&+&2\gamma_\mathrm{AFM}\boldsymbol{\Omega}-\underbrace{\chi\epsilon_\mathrm{AFM}\left[\ddot{\boldsymbol{\Omega}}+\dot{\boldsymbol{\Omega}}\times\boldsymbol{\Omega}\right]}_\mathrm{damping}+\frac{\gamma^2}{2\chi}\hat{\lambda}^{-1}\frac{\partial
 U_\mathrm{AFM}}{\partial\boldsymbol{\varphi}}=\nonumber\\
 &=& \underbrace{b_\mathrm{AFM}(\mathbf{j}\cdot\nabla)\boldsymbol{\Omega}}_{\mathrm{AST}}-\underbrace{
 c_\mathrm{AFM}\left[(\mathbf{j}\cdot\nabla)\dot{\boldsymbol{\Omega}}-\boldsymbol{\Omega}\times(\mathbf{j}\cdot\nabla)\boldsymbol{\Omega}\right]}_{\mathrm{NAST}},
 \end{eqnarray}
where we introduced the internal damping with factor
$2\gamma_\mathrm{AFM}$ calculated as a linewidth of AFMR,
$\hat{\lambda}^{-1}$ is the tensor inverse to $\hat{\lambda}$.
Mind, that the damping coefficient $\gamma_\mathrm{AFM}$ differs
from the analogous coefficient for FM system due to so-called
exchange enhancement: $\gamma_\mathrm{AFM}\propto
\gamma_\mathrm{FM}/\chi$, see, e.g., Ref.~\cite{gomo:2010}.

Equation (\ref{eq_dynamic_6}) includes three groups of terms that
stipulate from interaction between localized and free electrons.
The first group, with factor $\epsilon_\mathrm{AFM}$, is
independent on current. It accounts for additional damping related
with the itinerant electrons. To clarify this moment let us
consider the simple example of AFM rotation (or oscillation)
around a fixed axis $\mathbf{e}$. In this case
$\boldsymbol{\Omega}=\dot{\theta}\mathbf{e}$. In the absence of
current ($\mathbf{j}=0$) Eq.~(\ref{eq_dynamic_6}) takes a form:
\begin{equation}\label{eq_simple_dynamics_1}
  \ddot{\theta}+2\gamma_\mathrm{AFM}\dot{\theta}-\chi\epsilon_\mathrm{AFM}\dddot{\theta}+\frac{\gamma^2}{\chi}\frac{\partial U_\mathrm{AFM}}{\partial
  \theta}=0.
\end{equation}
For small oscillations with frequency $\Omega_\mathrm{AFMR}$ the
effective damping is renormalized as follows:
\begin{equation}\label{eq_renormalized damping}
  2\gamma_\mathrm{AFM}\quad\Rightarrow\quad2\gamma_\mathrm{AFM}+\chi\epsilon_\mathrm{AFM}\Omega^2_\mathrm{AFMR}.
\end{equation}
Analogous effect is observed in FMs, however, renormalization
(\ref{eq_renormalized damping}) in AFMs is frequency-dependent. It
should be mentioned that combination $\chi\Omega^2_\mathrm{AFMR}$
in AFMs is proportional to the magnetic anisotropy which is
usually small. In general case the additional damping results from
rotation around fixed axis (term $\ddot{\boldsymbol{\Omega}}$) and
from the axis rotation (term
$\dot{\boldsymbol{\Omega}}\times\boldsymbol{\Omega}$). However,
these contributions are important for fast modes only and could be
neglected for  $\Omega\leq\Omega_\mathrm{AFMR}$.

It worth to note that the above introduced damping ($\propto
\gamma_\mathrm{AFM}\boldsymbol{\Omega}$) models the ``viscous
resistance'' against the solid-like motion of localized magnetic
moments. In other words, we neglect the ``exchange damping'' that
hampers mutual rotation of the magnetic sublattices and growth of
$\mathbf{M}_\mathrm{AFM}$. The rather complicated problems of the
exchange damping are out of scope of this paper, however, within
the simplest phenomenological model (see,
e.g.\cite{Bar'yakhtar:1984E, Tserkovnyak:PhysRevLett.106.107206})
corresponding contribution into equation of motion is proportional
to high order time/space derivatives and thus has the same
structure as $\epsilon_\mathrm{AFM}$-term in (\ref{eq_dynamic_6}).

The second group, with the factor $b_\mathrm{AFM}$, is analogous
to (and its value coincides with) the AST in FMs
\cite{Bazaliy:PhysRevB.57.R3213, Zhang:PhysRevLett.93.127204}. The
value $b_\mathrm{AFM}\mathbf{j}$, which, in fact, is independent
on $sd$-exchange constant, can be interpreted as a relative
velocity of AFM texture with respect to steady current.  So, AST
produces a similar kinematic effect in FM and AFM textures.
However, in contrast to FMs, this term cannot be excluded from
Eqs.~(\ref{eq_dynamic_6}). To illustrate this fact, we again
consider rotation of AFM vectors around the fixed axis
$\mathbf{e}$ and take into account inhomogeneous exchange coupling
(constant $\alpha_\mathrm{inh}$) and possible nonlinearity of the
magnetic anisotropy (modeled with the potential
$U_\mathrm{an}(\theta)$):
\begin{equation}\label{eq_potential_energy}
  U_\mathrm{AFM}=\frac{1}{2}\alpha_\mathrm{inh}\left(\nabla\theta\right)^2+U_\mathrm{an}(\theta).
\end{equation}

Then, in neglection of dissipation (that includes damping and
NAST) Eq.~(\ref{eq_simple_dynamics_1}) takes a form:
\begin{equation}\label{eq_simple_dynamics_2}
  \ddot{\theta}-{v_\mathrm{mag}^2}\Delta \theta+
  \frac{\gamma^2}{\chi}\frac{dU_\mathrm{an}}{d
  \theta}=b_\mathrm{AFM}(\mathbf{j}\cdot\nabla)\dot{\theta},
\end{equation}
where $v_\mathrm{mag}\equiv\gamma\sqrt{\alpha_\mathrm{inh}/\chi}$
is the minimal phase velocity of magnons \cite{Ivanov:1983E}. In
the absence of current and anisotropy,
Eq.~(\ref{eq_simple_dynamics_2}) describes the Lorentz-invariant
dynamics (as was noticed for the first time in
Ref.~\cite{Bar_jun:1983} for the collinear AFM). As a result, AST
redefines the characteristic lengthscale of stationary moving
nonlinear waves. Really, the solution of
Eq.~(\ref{eq_simple_dynamics_2}),
$\theta_\mathrm{stab}(x-v_\mathrm{DW}t)$, that describes a
solitary wave moving with the constant velocity $v_\mathrm{DW}$
along the current direction ($\mathbf{j}\|x$), should satisfy the
following equation
\begin{equation}\label{eq_simple_dynamics_3}
  \left(v_\mathrm{mag}^2-b_\mathrm{AFM}jv_\mathrm{DW}-{v_\mathrm{DW}^2}\right)\frac{d^2\theta}{dx^2} =
  \frac{\gamma^2}{\chi}\frac{dU_\mathrm{an}}{d
  \theta}.
\end{equation}
Thus, the characteristic scale of inhomogeneity, $x_\mathrm{DW}$,
is current-dependent:
\begin{equation}\label{eq_characteristic lengthscale}
  x_\mathrm{DW}\quad\Rightarrow\quad
  \frac{x_\mathrm{DW}}{\sqrt{1-b_\mathrm{AFM}jv_\mathrm{DW}/v_\mathrm{mag}-{v_\mathrm{DW}^2/v_\mathrm{mag}^2}}}.
\end{equation}

At last, the third group in Eq.~(\ref{eq_dynamic_6}) is
responsible for NAST which can compensate the internal losses and
provide, in combination with the external magnetic field, a stable
motion of the AFM DW. For illustration we rewrite
Eq.~(\ref{eq_simple_dynamics_3}) for one dimensional
inhomogeneuity in the current direction ($\mathbf{j}\|x$) with
account of dissipation, NAST, and the constant external magnetic
field $H$ that sets the DW into motion:
\begin{equation}\label{eq_simple_dynamics_4}
  \ddot{\theta}-b_\mathrm{AFM}j\frac{\partial\dot{\theta}}{\partial x}-{v_\mathrm{mag}^2}\frac{\partial^2 \theta}{\partial x^2}+
  \frac{\gamma^2}{\chi}\frac{dU_\mathrm{an}}{d
  \theta}=-2\gamma_\mathrm{AFM}\dot{\theta}+
 c_\mathrm{AFM}j\frac{\partial\ddot{\theta}}{\partial x}+\frac{\gamma^2}{2}H^2\sin2\theta.
\end{equation}
Equation (\ref{eq_simple_dynamics_4}) can be treated as the
Lagrange equation in the presence of the external dissipative
forces with the Lagrange function
\begin{equation}\label{eq_Lagrange_simple}
 \mathcal{L}= \frac{\chi}{2\gamma^2}\left(\dot{\theta}-b_\mathrm{AFM}j\frac{\partial\theta}{\partial
  x}\right)^2-\frac{\chi}{2\gamma^2}\left(v_\mathrm{mag}^2-b^2_\mathrm{AF}j^2\right)\left(\frac{\partial\theta}{\partial
  x}\right)^2-U_\mathrm{an}(\theta)-\frac{\chi H^2}{4}\sin^2\theta%проверить поле
\end{equation}
and dissipation Rayleigh function
\begin{equation}\label{eq_Rayleigh_simple}
 \mathcal{R}=\frac{\chi}{\gamma^2}\left(\gamma_\mathrm{AFM}\dot{\theta}^2-
 c_\mathrm{AFM}j\dot{\theta}\frac{\partial^2}{\partial t^2}\frac{\partial{\theta}}{\partial
 x}\right),
\end{equation}
where generalized thermodynamic forces $\partial \theta/\partial
x$ are fixed (i.e. variation of the Rayleigh function should be
taken with respect to generalized velocities $\dot{\theta}$ only).

In the absence of current Eq.~(\ref{eq_simple_dynamics_4}) has a
standard DW solution
\begin{equation}\label{eq_dw_solution}
   \theta_\mathrm{stab}(x-v_\mathrm{DW}t)=\frac{1}{2}\arctan\left[\sinh\left(\frac{x-v_\mathrm{DW}t}{2x_\mathrm{DW}}\right)\right]
\end{equation}
The velocity of stable motion $v_\mathrm{DW}\equiv v_H=(\gamma H)^2 x_\mathrm{DW}/\gamma_\mathrm{AFM}=v_\mathrm{mag}(\gamma H)^2 /(\gamma_\mathrm{AFM}\Omega_\mathrm{AFMR})$ is defined from a balance %%$\omega_\mathrm{AFMR}=s/x_\mathrm{DW}$
between the energy losses (given by dissipation function) and
field-induced ponderomotive force:
\begin{equation}\label{eq_energy_losses}
-\int_{-\infty}^{\infty}\dot{\theta}\frac{\partial\mathcal{R}}{\partial\dot{\theta}}dx=\chi
H^2\int_{-\infty}^{\infty}\dot{\theta}\sin(2\theta) dx
\end{equation}

In the presence of current the stationary (nondissipative)
solution (\ref{eq_dw_solution}) is substantiated by average
compensation of losses. Substituting (\ref{eq_dw_solution}) into
Eq.~(\ref{eq_energy_losses}) we get the following expression for
the nonzero velocity $v^{(\mathrm{stab})}_\mathrm{DW}\ll
v_\mathrm{mag}$ (when the velocity dependence
(\ref{eq_characteristic lengthscale}) of $x_\mathrm{DW}$ could be
neglected):

\begin{equation}\label{eq_velocity_stable}
v^{(\mathrm{stab})}_\mathrm{DW}(j)=\frac{\gamma_\mathrm{AFM}v_\mathrm{mag}^2}{3c_\mathrm{AFM}j\Omega^2_\mathrm{AFMR}}\left(1-\sqrt{1-\frac{6c_\mathrm{AFM}j\Omega^2_\mathrm{AFMR}v_H}{\gamma_\mathrm{AFM}v_\mathrm{mag}^2}}\right)\approx
v_H\left(1+\frac{3c_\mathrm{AFM}j\Omega^2_\mathrm{AFMR}v_H}{\gamma_\mathrm{AFM}v_\mathrm{mag}^2}\right).
\end{equation}
Equation (\ref{eq_velocity_stable}) shows that the current,
depending on the direction, can either accelerate or slow down the
velocity of the already moving DW. Thus, the field-induced
velocity of stationary motion can be increased due to partial
compensation of damping. In principle, the current itself (i.e. in
the absence of the external magnetic field)  can also set the DW
into motion. However, transition from rest to motion goes through
destabilization of certain excitations and needs special treatment
which is out of scope of this paper.

It can be concluded that in AFM with the first type of
$sd$-exchange ($J_\mathrm{sd}^\mathrm{AFM}=0$ in Eq.~(\ref{eq_sd
interaction_AFM})) the current-induced effects could be observed
only for moving textures with $\boldsymbol{\Omega}\ne0$. However,
these effects could be pronounced and could be used for additional
control of the DW motion.

\section{Effect of antiferromagnetic-like $sd$-exchange:  artificial antiferromagnet}
\begin{figure}[hp]
  \includegraphics[width=0.6\textwidth]{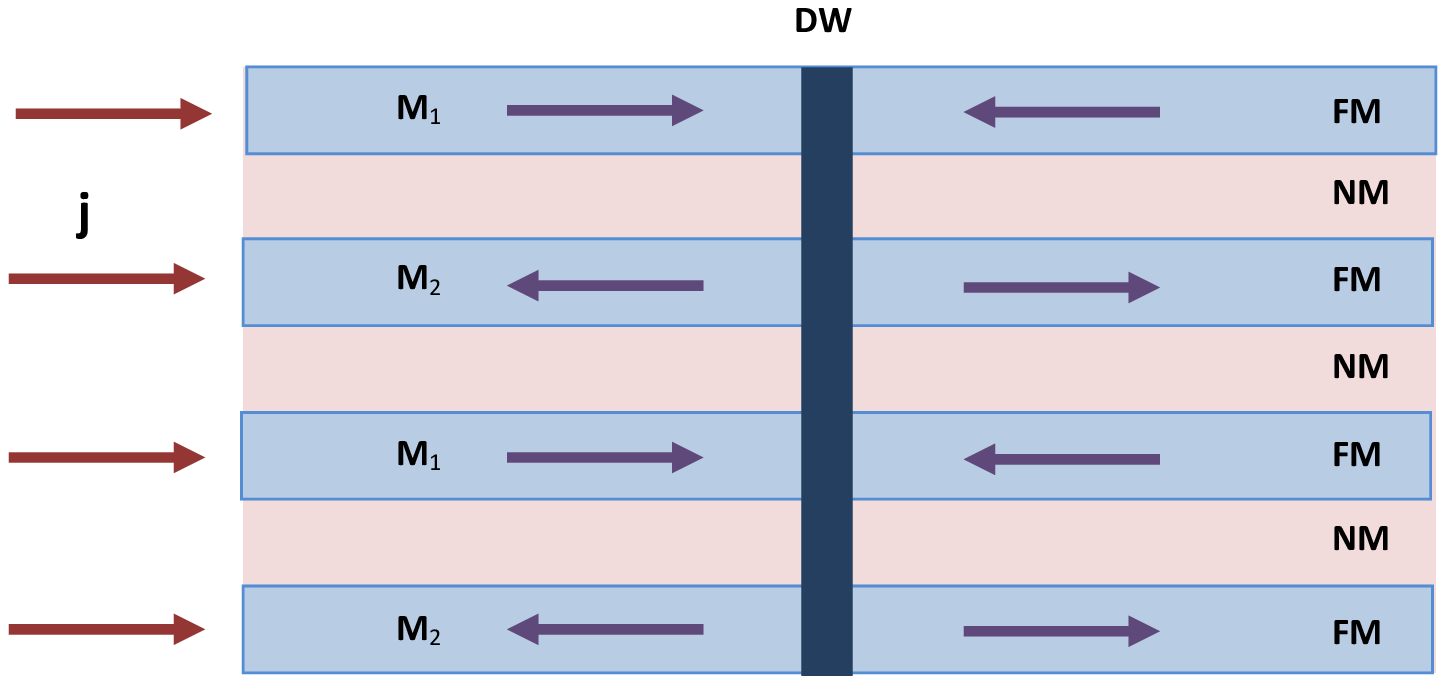}\\ %%for preprint
  \caption{(Color online) Artificial AFM with the 180$^\circ$ DW. Distribution of
  magnetization vectors $\mathbf{M}_1$, $\mathbf{M}_2$, (thick
  arrows) within each of FM sublayers is inhomogeneous, however,
  due to strong interlayer coupling rotation of magnetization in
  different layers is coherent forming the global DW
  (vertical stripe). Nonpolarized current $\mathbf{j}$ flows
  parallel to the easy axis of the layer.
  }\label{fig_multilayered_structure}
\end{figure}

In this section we consider an artificial system consisting of
$N_\mathrm{layer}$ well separated antiferromagnetically coupled FM
layers (see, e.g., \cite{Fullerton:PhysRevLett.91.197203,
Aliev:2009PhRvB..79m4423H, sbiaa:242502}). Experiments show that
even for $N_\mathrm{layer}\propto 10$ such multilayers demonstrate
the characteristic features of AFM:  small magnetic susceptibility
that points to strong exchange coupling between the layers
\cite{hashimoto:032511}, spin-flop transition in the external
magnetic field \cite{xiao:17A325}, DWs that penetrate all the
layers \cite{Hellwig:NatM_2003, Bogdanov:Kiselev20101340} (see
Fig.\ref{fig_multilayered_structure}).

  In the case of
current-in-plane configuration each layer -- macroscopic
sublattice, -- can polarize the current along the local
magnetization vector $\mathbf{M}_k$ ($k=1,\ldots,
N_\mathrm{layer}$) thus producing a modulation of spin
polarization. So, the energy density of $sd$-exchange is described
with the expression (cp.(\ref{eq_sd interaction_AFM}))
 \begin{equation}\label{eq_sd_artificial_1}%%M_0\rightarrow --Ms
    \mathcal{H}_\mathrm{sd}=-\frac{J_\mathrm{sd}}{M_{s}N_\mathrm{layer}}\sum_{k=1}^{N_\mathrm{layer}}\mathbf{M}_k\mathbf{m}_k,
\end{equation}
where $\mathbf{m}_k$ is electron magnetization density of the
$k$-th layer and $M_{s}=|\mathbf{M}_k|$. For a collinear structure
consisting of two equivalent magnetic sublattices $\mathbf{M}_1$
and $\mathbf{M}_2$ Eq.~(\ref{eq_sd_artificial_1}) can be presented
in the form analogous to (\ref{eq_sd interaction_AFM}) with
$J_\mathrm{sd}^\mathrm{FM}=J_\mathrm{sd}^\mathrm{AFM}=J_\mathrm{sd}$
\begin{equation}\label{eq_sd_artificial_2}
    \mathcal{H}_\mathrm{sd}=-\frac{J_\mathrm{sd}}{4M_{s}}\left(\mathbf{m}\mathbf{M}+\mathbf{l}\mathbf{L}\right),
\end{equation}
where $\mathbf{L}=\mathbf{M}_1-\mathbf{M}_2$ is AFM vector, $\mathbf{M}=\mathbf{M}_1+\mathbf{M}_2$ is macroscopic magnetization of localized spins and $\mathbf{l}$ and $\mathbf{m}$ are analogous combinations for free electrons.

If we neglect small perpendicular spin transfer between the
layers\footnote{~We cannot exclude the perpendicular motion of
carriers between the layers which provides exchange interaction
through RKKY mechanism. However, for CIP (i.e. current-in-plane)
configuration the main contribution into nonequilibrium spin flux
occurs from the in-plane component of electron velocity.}, the
current-induced dynamics of sublattice magnetization
$\mathbf{M}_k$  within the sublayer ($k=1,2$) can be described
with the standard equation for FM
\cite{Zhang:PhysRevLett.93.127204,Sun:2011EPJB_79_449S}:
\begin{equation}\label{eq_LLGS_continuous}%знаки по Цангу
    \dot{\mathbf{M}}_k=-\gamma\left[\mathbf{M}_k\times\mathbf{H}^\mathrm{eff}_k\right]+\frac{\alpha_G}{M_{s}}\left[\mathbf{M}_k\times\dot{\mathbf{M}_k}\right]+b(\mathbf{j}\cdot\nabla)\mathbf{M}_k-\frac{c}{M_{s}}\left[\mathbf{M}_k\times(\mathbf{j}\cdot\nabla)\mathbf{M}_k\right], \end{equation}
where $\mathbf{H}^\mathrm{eff}_k\equiv-(\partial U/\partial
\mathbf{M}_k)$ ($U$ is the density of magnetic energy), $\alpha_G$
is the constant of Gilbert damping (inversely proportional to the
FMR quality factor), coefficients $b=\mu_BP/eM_0(1+\xi^2)$ and
$c=b\xi$ (where $\xi\equiv\hbar/(\tau_\mathrm{sf}J_\mathrm{sd})\ll
1$) describe AST and NAST within each FM layer, correspondingly.
The sign of the first term in the r.h.s. of
Eq.~(\ref{eq_LLGS_continuous}) is related with the chosen positive
sign of gyromagnetic ratio, $\gamma>0$.

In contrast to a single, isolated FM layer, the effective field
$\mathbf{H}^\mathrm{eff}_k$ in artificial magnet includes
contribution from the exchanged coupling. In particular, in the
simplest case of a collinear AFM:
\begin{equation}\label{eq_energy_magnetic}
    U=U_\mathrm{AFM}(\mathbf{L},\nabla\mathbf{L})+\frac{H_\mathrm{ex}}{4M_{s}}\mathbf{M}^2,
\end{equation}
where the energy density
$U_\mathrm{AFM}(\mathbf{L},\nabla\mathbf{L})$ includes the
magnetic anisotropy and contribution from inhomogeneous exchange
(cp. with (\ref{eq_potential_energy})), $H_\mathrm{ex}$, as above,
is the constant of the exchange coupling equal to spin-flip field.

The set of Eqs.~(\ref{eq_LLGS_continuous}) can be simplified in
approximation of rather strong exchange coupling between
sublattices-sublayers, i.e., when the characteristic values of the
external fields (including current-induced effects) are much
smaller than $H_\mathrm{ex}$ and thus $|\mathbf{M}|\ll
|\mathbf{L}|$.  First, we rewrite equations
(\ref{eq_LLGS_continuous}) in terms of magnetization,
$\mathbf{M}$, and AFM vector, $\mathbf{L}$, as follows:
\begin{eqnarray}\label{eq_SAF_dynamics}%Знак затухания!\gamma<0
\dot{\mathbf{M}}&=&-\gamma\mathbf{L}\times\mathbf{H}_\mathrm{L}+\frac{\alpha_G}{2M_{s}}\left[\mathbf{L}\times\dot{\mathbf{L}}\right]+b(\mathbf{j}\cdot\nabla)\mathbf{M}-\frac{c}{2M_{s}}\left[\mathbf{L}\times(\mathbf{j}\cdot\nabla)\mathbf{L}\right]\\\nonumber
\dot{\mathbf{L}}&=&\frac{\gamma
H_\mathrm{ex}}{2M_{s}}\mathbf{L}\times\mathbf{M}+b(\mathbf{j}\cdot\nabla)\mathbf{L},
\end{eqnarray}
where the effective field $\mathbf{H}_\mathrm{L}\equiv-(\delta
U_\mathrm{AFM}/\delta \mathbf{L})$.

Then, following the standard approach proposed by  Bar'yakhtar and
 Ivanov \cite{Bar-june:1979E, Turov:1998E}, we exclude
magnetization $\mathbf{M}$ from the second of
Eqs.~(\ref{eq_SAF_dynamics}) thus obtaining the self-consistent
dynamic equation for AFM vector in the presence of current:
\begin{eqnarray}\label{eq_SAF_dynamics_2}%%to check coefficients!
\mathbf{L}&\times&\left[\ddot{\mathbf{L}}-2\gamma^2 M_{s}
H_\mathrm{ex} \mathbf{H}_\mathrm{L}+\gamma
H_\mathrm{ex}\alpha_G\dot{\mathbf{L}}-2b(\mathbf{j}\cdot\nabla)\dot{\mathbf{L}}\right.\\
&-&\left.b\left(\frac{d\mathbf{j}}{dt}\cdot\nabla\right)\mathbf{L}+b^2(\mathbf{j}\cdot\nabla)^2\mathbf{L}-\gamma
H_\mathrm{ex}c(\mathbf{j}\cdot\nabla)\mathbf{L}\right]=0.\nonumber%дописать
\end{eqnarray}
Corresponding Lagrange and Rayleigh functions  are
\begin{equation}\label{eq_Lagrange-SAF}%%coefs!
 \mathcal{L}=\frac{1}{4\gamma^2H_\mathrm{ex}M_{s}}\left(\dot{\mathbf{L}}-b(\mathbf{j}\cdot\nabla)\mathbf{L}\right)^2- U_\mathrm{AFM},%дописать
\end{equation}
and
\begin{equation}\label{eq_Rayleigh-SAF}
 \mathcal{R}=\frac{\alpha_G}{4\gamma M_{s}}\dot{\mathbf{L}}^2-\frac{c}{2\gamma M_{s}}\dot{\mathbf{L}}(\mathbf{j}\cdot\nabla)\mathbf{L}.
\end{equation}
As above, the Railegh function (\ref{eq_Rayleigh-SAF}) is
considered at fixed generalized thermodynamic forces
$\nabla\mathbf{L}$.

To illustrate  the peculiar features of the current-induced
phenomena in artificial, or synthetic antiferromagnets (SyAFMs),
we consider the simplest example of an easy-axis AFM whose
dynamics can be described with the single variable $\theta$ (angle
between $\mathbf{L}$ and easy axis) and the density of the direct
current $\mathbf{j}\|z$. Then, dynamic
Eq.~(\ref{eq_SAF_dynamics_2}), Lagrange(\ref{eq_Lagrange-SAF}),
and Rayleigh functions (\ref{eq_Rayleigh-SAF}) take the following
form (cp. with their counterparts (\ref{eq_simple_dynamics_4}),
(\ref{eq_Lagrange_simple}), (\ref{eq_Rayleigh_simple})):
\begin{equation}\label{eq_SyAF_dynamics_4}
  \ddot{\theta}-bj\frac{\partial\dot{\theta}}{\partial z}-({v_\mathrm{mag}^2-b^2j^2})\frac{\partial^2 \theta}{\partial z^2}-v_\mathrm{mag}^2\Delta_\perp\theta+
  \frac{\gamma^2}{\chi}\frac{dU_\mathrm{an}}{d
  \theta}=-2\gamma_\mathrm{AFM}\dot{\theta}-
 \frac{\gamma cjH_\mathrm{ex}}{2}\frac{\partial\theta}{\partial z},
\end{equation}
where $U_\mathrm{an}$ is the density of the magnetic anisotropy
energy, $\Delta_\perp$ is two-dimensional Laplace operator in $xy$
plane,
\begin{equation}\label{eq_Lagrange_SyAF_simple}%%to check!
 \mathcal{L}= \frac{\chi}{2\gamma^2}\left(\dot{\theta}-bj\frac{\partial\theta}{\partial
  z}\right)^2-\frac{\chi}{2\gamma^2}\left(v_\mathrm{mag}^2-b^2j^2\right)\left(\frac{\partial\theta}{\partial
  z}\right)^2-U_\mathrm{an}(\theta),%проверить поле
\end{equation}
and
\begin{equation}\label{eq_Rayleigh_SyAF_simple}%%to check!
 \mathcal{R}=\frac{\chi}{\gamma^2}\left(\gamma_\mathrm{AFM}\dot{\theta}^2-
 \gamma cjH_\mathrm{ex}\dot{\theta}\frac{\partial\theta}{\partial
 z}\right),
\end{equation}
where $2\gamma_\mathrm{AFM}=\gamma\alpha_GH_\mathrm{ex}$ and
$\chi=2M_{s}/H_\mathrm{ex}$, as above.

Comparison of Eqs.~(\ref{eq_simple_dynamics_4}) and
(\ref{eq_SyAF_dynamics_4}) shows that AST described with the
constant $b=b_\mathrm{AFM}$ has \emph{exactly} the same form in
FM, AFM and SyAFM. In any type of AFM, regardless of type of
$sd$-exchange, AST results in kinematic effects and reveals itself
in the renormalization of the DW width (see
Eq.~(\ref{eq_characteristic lengthscale})).

The main difference between AFMs with FM- and AFM-like
$sd$-exchange shows in NAST (term with $c$). First of all,
corresponding constants of NAST, $c_\mathrm{AFM}$ (see
(\ref{eq_constants})) and $c$ in (\ref{eq_LLGS_continuous}) have
different dimensionality and different microscopic origin. Roughly
speaking, in artificial systems  (and in FMs) the NAST arises
mainly from the second term in the l.h.s. of general relation
(\ref{eq_nonequilibrium_magnetization}), i.e. from rotation of
free electron spins around the local magnetization. In contrast,
this process is neglected in AFMs with the FM-like $sd$-exchange,
as was already discussed above.

Second, while in ``natural'' AFMs the nonadiabatic spin torque has
pure dynamic origin, i.e. is proportional to time derivatives of
$\theta$, in artificial AFMs the NAST appears in the region of
inhomogeneuity and is proportional to space derivative of
$\theta$.

Third, in analogy with FMs, the NAST in SyAFMs can compensate the
internal losses and ensure steady motion of the DW. Really, a
soliton-like solution $\theta_\mathrm{stab}(z-v_\mathrm{DW}t)$
makes vanish the r.h.s. of Eq.~(\ref{eq_SyAF_dynamics_4}) if
\begin{equation}\label{eq_critical_velocity}% to write further
  v_\mathrm{DW}=\frac{cj}{\alpha_G}=\frac{\gamma cjH_\mathrm{ex}}{2\gamma_\mathrm{AFM}}.
\end{equation}

Analysis of Eq.~(\ref{eq_critical_velocity}) shows that the
velocity of steady motion in SyAFM, in analogy with FMs and in
contrast to ``natural'' AFMs (cp. with
(\ref{eq_velocity_stable})), is proportional to the current
density $j$ and is defined by the balance of damping (constant
$\alpha_G$) and NAST (constant $c$).
Expression~(\ref{eq_critical_velocity}) can also be interpreted in
other aspect. Really, the FMR quality factor  (inversely
proportional to $\alpha_G$) is usually greater than the AFMR
quality factor. This difference stems from the exchange
enhancement of damping coefficient: $\gamma_{AFM}\propto
\alpha_GH_\mathrm{ex}$, while $\Omega_\mathrm{AFMR}\propto
\sqrt{H_\mathrm{an}H_\mathrm{ex}}$, so, the quality factor
$\Omega_\mathrm{AFMR}/\gamma_{AFM}\propto
\sqrt{H_\mathrm{an}/H_\mathrm{ex}}/\alpha_G\ll 1/\alpha_G$.
However, in the case of SyAFM both NAST and damping are enhanced
in a similar way, as seen from the second equality in
(\ref{eq_critical_velocity}). So, the velocity of steady motion in
SyAFM has the same value as the velocity of current-induced DW
motion in the FM constituents of this artificial structure. It
should be also mentioned that the experiments
\cite{Aliev:2009PhRvB..79m4423H} point to current-induced DW
motion in the SyAFM consisting of Fe/Cr multilayers.

\begin{figure}[htbp]
    \centering
            \includegraphics[width=0.6\textwidth]{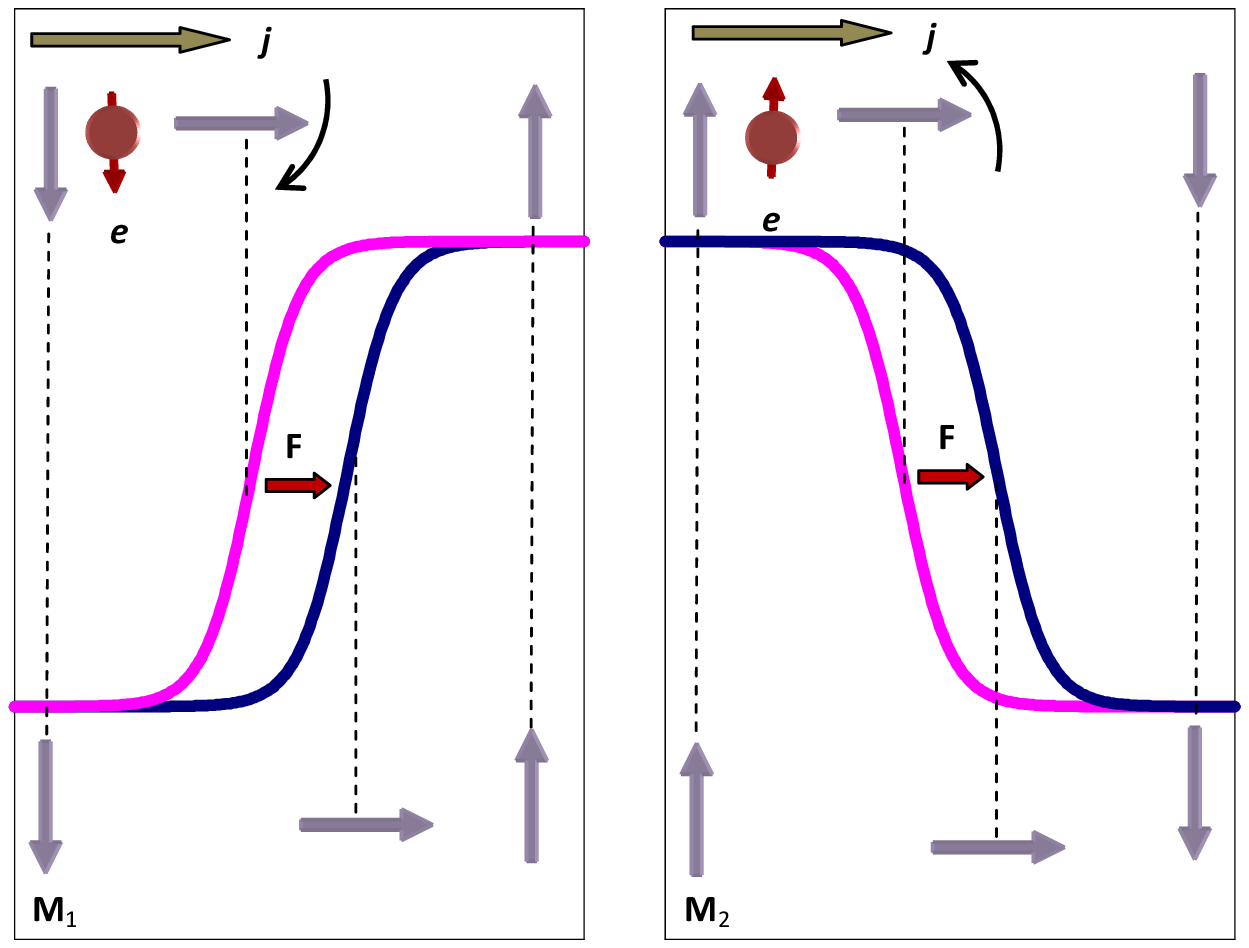}%%for preprint
  \caption{(Color online) Current-induced motion of 180$^\circ$ DW in artificial AFM.
  Lines show the magnetization profiles $\mathbf{M}_1(z)$,
  $\mathbf{M}_2(z)$ within each of FM sublayers. Arrows in upper
  (lower) row schematically  show the orientation of magnetization
  in different points (dashed lines) before (after) interaction
  with the free electron $e$. In both cases the current pushes the
  DW in the same direction and thus produces the same
  ponderomotive force $\mathbf{F}$ in each
  sublayer.}\label{fig_ponderomotive_Syaf}
\end{figure}

At last, we would like to note that according to
Eqs.~(\ref{eq_SAF_dynamics_2}) and~(\ref{eq_SyAF_dynamics_4}) the
current can set into motion the 180$^\circ$ DW in SyAFM, while the
constant magnetic field can not. Really, the 180$^\circ$ DW
separates the translation domains (with $\mathbf{L}$ and
$-\mathbf{L}$) which, due to quadratic dependence of Zeeman energy
($\propto (\mathbf{H}\times\mathbf{L})^2$), keep equivalence in
the constant magnetic field. In contrast, the staggered
polarization of current in the SyAFM ensures the same direction of
the ponderomotive force in each FM sublayer (see
Fig.\ref{fig_ponderomotive_Syaf}). Thus, the current-induced
motion of the DW in SyAFM is equivalent to the motion of
$N_\mathrm{layer}$ FM domain walls synchronized due to AFM
exchange coupling between the layers.

\section{Discussion: comparison of different models}
Above we obtained two forms of dynamics equations for AFM texture
in the presence of current, starting from different forms of
$sd$-coupling. Thus, Eqs.~(\ref{eq_dynamic_6}) and
(\ref{eq_SAF_dynamics_2}) constitute the most essential result of
our paper.

Analysis of these equations shows that the continuous AFM systems,
in analogy with FM textures, should demonstrate the
current-induced dynamics. However, peculiarities of
current-induced phenomena strongly depend upon the details of band
structure and, particularly, of coupling between the
spin-contributing localized electrons and carriers. We expect that
in the alloys with well separated $d$- and $s$-bands (like FeMn
\cite{Nakamura:PhysRevB.67.014405} or IrMn) the $sd$-exchange has
the FM-like structure (\ref{eq_Heisenberg-like term}) and
current-induced dynamics is modeled with Eq.~(\ref{eq_dynamic_6}).
Current-induced effects in this case have the dynamic origin and
could be observed in the moving textures or along with field- or
temperature-driven oscillations of AFM vectors.

The second, AFM-like type of $sd$-exchange can be expected in
artificial AFMs, where the staggered spin modulation of free
electrons is imposed by the geometry of superstructure or in the
itinerant AFMs like Cr (interconnection between the transport
properties and AFM order in Cr was recently observed in
Ref.\onlinecite{Kummamuru:2011}). However, applicability of
$sd$-exchange model in the latter case needs further
investigation.

It is instructive to compare the Eqs.~(\ref{eq_dynamic_6}) and
(\ref{eq_SAF_dynamics_2}) with the models proposed by Hals et al.
\cite{Tserkovnyak:PhysRevLett.106.107206} and  Swaving and Duine
 \cite{Duine:PhysRevB.83.054428,
Duine:0953-8984-24-2-024223}. For this purpose we reproduce four
known dynamic equations for the collinear AFM along with the short
description. For the sake of clarity we change some of the
original author's notations, so that AFM order parameter (AFM, or
N\'{e}el, vector) is denoted as $\mathbf{L}$, exchange constant as
$J_\mathrm{ex}\equiv 2H_\mathrm{ex}M_{s}$, etc.

\begin{enumerate}
  \item Equation (8) of Ref.~\cite{Tserkovnyak:PhysRevLett.106.107206}:
  \begin{eqnarray}\label{eq_Tserkovnyak}
    \ddot{\mathbf{L}}&=&-\widetilde{\gamma} \mathbf{L}\times \dot{\mathbf{H}}+\widetilde{\gamma}
    G_1\dot{\mathbf{H}}_L+\widetilde{\gamma}(\eta+G_1\beta)\underline{\left(\frac{d\mathbf{j}}{dt}\cdot\nabla\right)\mathbf{L}}\nonumber\\
    &+&2\widetilde{\gamma}H_\mathrm{ex}M_{s}\left[\gamma\mathbf{H}_L-\alpha_G\dot{\mathbf{L}}+\underline{\gamma\beta(\mathbf{j}\cdot\nabla)\mathbf{L}}\right],
\end{eqnarray}
where $G_1$ is the exchange damping parameter (omitted in our
model), $\widetilde{\gamma}=\gamma/(1+G_1\alpha_G)$, $\eta$ and
$\beta$ parametrize AST and NAST, respectively. This equation is
derived from the general thermodynamic principles starting from
the hypothesis of spin pumping effect in AFMs. In other words, the
main assumption of the model is that the spins of carriers flowing
through AFM layer aquire the same magnetic ordering (staggered
magnetization) as the localized moments. This picture is
consistent with the model of artificial AFM considered in the
present paper and thus, Eq.~(\ref{eq_Tserkovnyak}) should be
compared with Eq.~(\ref{eq_SAF_dynamics_2}) derived in assumption
of AFM-like $sd$-exchange:
  \begin{eqnarray}\label{eq_gomo_1}
  \ddot{\mathbf{L}}&=&2\gamma^2 H_\mathrm{ex}M_{s}
\mathbf{H}_\mathrm{L}-\gamma\alpha_GH_\mathrm{ex}\dot{\mathbf{L}}+\underbrace{2b(\mathbf{j}\cdot\nabla)\dot{\mathbf{L}}-b^2(\mathbf{j}\cdot\nabla)^2\mathbf{L}}_\mathrm{2nd}\nonumber\\
&+&\underline{b\left(\frac{d\mathbf{j}}{dt}\cdot\nabla\right)\mathbf{L}}+\underline{\gamma
cH_\mathrm{ex}(\mathbf{j}\cdot\nabla)\mathbf{L}}.
\end{eqnarray}
To simplify the analysis, the similar current-induced terms in
both equations are underlined. The terms labeled as ``2nd'' in our
model are of the second order of value in $\mathbf{j}$,
$\dot{\mathbf{L}}$ and were omitted as small in
Ref.~\cite{Tserkovnyak:PhysRevLett.106.107206}. Thus, we can
conclude that both models (AFM spin pumping and AFM-like
$sd$-exchange models) predict the same dynamics and both work well
for artificial AFMs.

\item Equation (14) of Ref.~\cite{Duine:PhysRevB.83.054428} or (3) of
Ref.~\cite{Duine:0953-8984-24-2-024223}:
   \begin{eqnarray}\label{eq_swaving_1}
    \mathbf{L}&\times & \left(\ddot{\mathbf{L}}+\gamma_D\dot{\mathbf{L}}-v_\mathrm{mag}^2\nabla^2\mathbf{L}\right)+\frac{\partial}{\partial t}\left[\mathbf{L}\times(\mathbf{v}\cdot\nabla)\mathbf{L}\right]=\nonumber\\
&=&-\left(\mathbf{L}\cdot\dot{\mathbf{L}}\times(\mathbf{v}\cdot\nabla)\mathbf{L}\right)\mathbf{L}-v_\mathrm{mag}(\mathbf{r}\cdot\nabla)(\mathbf{v}\cdot\nabla)\mathbf{L}
\end{eqnarray}
where $\mathbf{r}=(111)^T$, $\gamma_D$ is the damping parameter.
This equation is derived for the structure in which the sublattice
magnetization rotates from site to site, i.e. sublattice
magnetizations $\mathbf{M}_1$, $\mathbf{M}_2$ are tilted with
respect each other even in the absence of field and
current\footnote{~In the collinear, (compensated) AFMs the space
inhomogeneuity of AFM vector is usually attributed to the
different \emph{physically} small volumes much greater than the
unit cell. Thus, each physical ``space point'' keeps the symmetry
properties of homogeneous AFM structure, in particular,
permutation symmetry for magnetic sublattices. In contrast, in
Refs.~\onlinecite{Duine:PhysRevB.83.054428,
Duine:0953-8984-24-2-024223} the space inhomogeneuity is
attributed to the lattice sites and thus concerns incommensurate
magnetic structures like spirals. In the last case permutation
symmetry is lost and AFM vector
$\mathbf{L}(\ne\mathbf{M}_1-\mathbf{M}_2$) is defined as order
parameter which belongs to irreducible representation of
Shubnikov's group with $\mathbf{k}\ne0$.}. This gives rise to a
Lifshitz-like contribution $\partial \mathbf{L}/\partial z$ to
magnetization (see Eq.~(12) of
Ref.~\onlinecite{Duine:PhysRevB.83.054428}) typical for
noncentrosymmetric incommensurate structures.  The current-induced
effects in this model arise from the noncompensated \emph{static}
magnetization omitted in both our models and in
Ref.~\cite{Tserkovnyak:PhysRevLett.106.107206}. It should be also
stressed that l.h.s. of Eq.~(\ref{eq_swaving_1}) should be
orthogonal to $\mathbf{L}$ and this condition imposes additional
limitations on the type of space inhomogeneuity.

 \item Equation (\ref{eq_dynamic_6}) derived from spin
  conservation principle assuming FM-like $sd$-exchange
  and adapted for the collinear AFM with two magnetic sublattices
  by substitution $\boldsymbol{\Omega} \rightarrow
  \mathbf{L}\times \dot{\mathbf{L}}$ and due account of orthogonality condition $(\mathbf{M}_\mathrm{AFM}\cdot\mathbf{L})=0$:
\begin{eqnarray}\label{eq_natural_collinear_AFM}%dopisat'
\mathbf{L} &\times&\left\{\ddot{\mathbf{L}}-2\gamma^2
H_\mathrm{ex}M_{s}
\mathbf{H}_\mathrm{L}+\gamma\alpha_GH_\mathrm{ex}\dot{\mathbf{L}}\right\}\nonumber\\
&=&b_\mathrm{AFM}\left\{(\mathbf{j}\cdot\nabla)\left(\mathbf{L}\times\dot{\mathbf{L}}\right)-\mathbf{L}\left[\mathbf{L}\cdot(\mathbf{j}\cdot\nabla)\mathbf{L}\times\dot{\mathbf{L}}\right]/4M_{s}^2\right\}\\
&+&(c_\mathrm{AFM}/4M_{s}^2)\left\{2\dot{\mathbf{L}}[\dot{\mathbf{L}}\cdot\mathbf{L}\times(\mathbf{j}\cdot\nabla)\mathbf{L}]+4M_{s}^2\ddot{\mathbf{L}}\times(\mathbf{j}\cdot\nabla)\mathbf{L}+\mathbf{L}[\ddot{\mathbf{L}}\cdot\mathbf{L}\times(\mathbf{j}\cdot\nabla)\mathbf{L}]\right\}\nonumber.
\end{eqnarray}
Comparison of Eq.~(\ref{eq_natural_collinear_AFM}) with
Eqs.~(\ref{eq_Tserkovnyak}) and (\ref{eq_gomo_1}) shows that
current-induced terms and, consequently, current-induced phenomena
predicted within two models of $sd$-exchange are absolutely
different. This difference opens a way to elucidate the mechanism
of $sd$-coupling in AFM materials from the peculiarities of
current-induced dynamics.
\end{enumerate}

\section{Conclusions}
To summarize, we considered the current-induced dynamics for
different types of the continuous AFMs and obtained equations that
could be used for the analysis of the dDWs, droplets and other
soliton-like structures in AFMs in the presence of current. The
predicted qualitative difference in dynamics for FM- and AFM-like
types of $sd$-exchange opens a way for experimental investigation
of the carrier's role in AFM ordering of a certain material.
\acknowledgments The authors acknowledge the fruitful discussions
with Yu.~Gaididei and D.~Scheka. H.G. is grateful to F.~G.~Aliev
who attracted her attention to the problem of artificial
antiferromagnets. The work is performed under the program of
fundamental Research Department of Physics and Astronomy, National
Academy of Sciences of Ukraine, and supported in part by a grant
of Ministry of Education and Science of Ukraine.
%\bibliography{../../../tex/spin_laser_2009} %home
%\bibliography{../../../plenki/LVIV/spin_laser_2009} %work%bibliography style is included into description of aps style.
%\end{document}

%Merlin.mbs v4.21 2009-07-09.
%

\end{document}